\documentstyle[12pt]{article}
\title{Path Integral in Holomorphic Representation without
Gauge Fixation\thanks{JINR preprint E2-89-678, JINR, Dubna, 1989
(unpublished)}}
\author{\large Sergei V. SHABANOV \\[3mm]
\em Department of Theoretical Physics,  \\
\em University of Valencia, \\
\em Moliner 50, Burjassot (Valencia), E-46100, Spain\\
\em and\\
\em Laboratory of Theoretical Physics, JINR, Dubna, Russia}
\date{}
\begin{document}
\maketitle
\begin{abstract}
A method of path integral construction without gauge fixing
in the holomorphic representation is proposed for finite-dimensional
gauge models. This path integral determines a manifestly gauge-invariant
kernel of the evolution operator.
\end{abstract}
\def\tint{{\textstyle\int}}
\newcommand{\la}{\langle}
\newcommand{\ra}{\rangle}
\newcommand{\pl}{\partial}
\newcommand{\be}{\begin{equation}}
\newcommand{\ee}{\end{equation}}
\newcommand{\ba}{\begin{eqnarray}}
\newcommand{\ea}{\end{eqnarray}}
\def\R{\relax{\rm I\kern-.18em R}}
\def\1{\relax{\rm 1\kern-.27em I}}
\newcommand{\Z}{Z\!\!\! Z}
\newcommand{\ph}{PS_{ph}}

{\bf 1}. It is well known that a gauge symmetry leads to constraints on
dynamical variables in the theory \cite{1}. Therefore, the evolution of
unphysical degrees of freedom should be given when working with gauge
theories, which implies gauge fixing.
Alternatively, one can go over
to gauge-invariant variables by means of an appropriate canonical
transformation. In the latter case constraints
turns into some of the new canonical momenta.
Gauge-invariant variables are, in general, described
by curvilinear coordinates, and their configuration
space differs from the Euclidean space \cite{2}, \cite{3}. In other words,
a physical coordinate may take its
value not on the whole real axis but only on
its part (a halfline or a segment). Moreover physical degrees of freedom
can have a phase space which differs
from a plane \cite{4}, \cite{5}. It leads to a modification of PI
\cite{5}, and as a result, the quasi-classical description is changed
\cite{6}.

According to the above remarks the following question can be raised: is
there any way to construct PI which does not require elimination of
unphysical degrees of freedom, and the evolution operator determined by
such PI would be manifestly gauge-invariant? It is shown below that
for finite-dimensional models with a gauge group (including the
Yang-Mills quantum mechanics \cite{7}) this question is not deprived
of sense, and the recipe of finding PI that involves no gauge condition
is proposed.

{\bf 2}. We shall explain the main idea of the note by a simple example
where there is only one physical degree of freedom. The Lagrangian of the
model is \cite{4}
\begin{equation}
L=(\dot{x}-y_aT^ax)^2/2-V(x^2)\ .
\label{1}
\end{equation}
Here an N-dimensional vector $x=(x_1,x_2,\ldots , x_N)$ and
$y_a(a=1,2,\ldots , N)$ play the role of dynamical variables of the theory,
$T^a$ are $N\times N$ antisymmetric matrices which are generators of the
group SO(N), $[T^a,T^b]=f_{abc}T^c$, $f_{abc}$ are structural constants of
SO(N), $(T^ax)_i=T^a_{ij}x_j$ and $V$ is a potential. Lagrangian
(\ref{1}) remains invariable with respect to gauge transformations
\begin{equation}
x\rightarrow \Omega x \Omega^T,\ \ y\rightarrow \Omega y\Omega ^T-
\Omega \pl _t\Omega ^T,\ \ y=y_aT^a\ ,
\label{2}
\end{equation}
where $\Omega =\exp \omega _a(t)T^a,\ \omega _a$ are arbitrary functions
of time, $\Omega ^T$ is the transposed matrix $\Omega $.

Going over to
the Hamiltonian formalism we find canonical momenta $\pi _a=
\pl L/\pl \dot{y}_a=0$ (primary constraints \cite{1}) and $p=\pl L/\pl
\dot{x}=\dot{x}-y_aT^ax$. The Hamiltonian is
\begin{equation}
H=p^2/2+V(x^2)-y_aG^a\ ,
\label{3}
\end{equation}
where $G_a=\{\pi _a,H\}=p_iT^a_{ij}x_j=0$ are secondary
constraints ($\{ , \}$ are Poisson brackets) which follow from the
consistency condition $\dot{\pi}_a=0$ \cite{1}. All constraints are
of the first class $\{G_a,G_b\}=f_{abc}G_c,\ \{G_a,H\}=-f_{abc}y_bG_c$.
Thereby the quantization of the theory is carried out by the change of
both the momenta and coordinates to operators with the commutation
relations $[x_j,p_k]=i\delta _{jk},\ [y_a,\pi _b]=i\delta _{ab}$, while
the constraints select physical states \cite{1}:
\begin{equation}
G_a|\psi_{ph}\ra =0,\ \ \pi _a|\psi _{ph}\ra =0\ .
\label{4}
\end{equation}
The second equality in (\ref{4}) means that wave functions do not depend
on $y_a$; so below we shall not take these degrees of freedom into
consideration. The first equation of (\ref{4}) can easily be solved in the
holomorphic representation.
We define the operators
\cite{8} $\hat{a}_j=(x_j+ip_j)/\sqrt{2}$ and the representation
$\hat{a}^+_j\psi
(a^*)=a^*_j\psi (a^*),\ \hat{a}_j\psi (a^*)=\pl /\pl a^*_j\psi (a^*)$. The
scalar product reads
\begin{equation}
\tint d^N(a^*,a)(\psi _1(a^*))^*\psi _2(a^*)=\la \psi _1|\psi
_2\ra\ ,
\label{5}
\end{equation}
where $d^N(a^*,a)=(2\pi i)^{-N}d^Na^*d^Na\exp (-a^*_ja_j)$. Any state in
the holomorphic representation is decomposed over the basis $\la
a^*|n_1,\ldots n_N\ra =\prod_{i=1}^{N}(a^*_i)^{n_i}/\sqrt{n_i!}$ here
$n_i=0,1,\ldots$. This basis is orthonormal with respect to the scalar
product (\ref{5}). The constraint operators become
$G_a=T^a_{ij}\hat{a}^+_i\hat{a}_j$. Note that here there is no
operator ordering problem as $T^a$ are antisymmetric matrices.

Clearly, the vacuum $\la a^*|0\ra =1$ satisfies (\ref{4}), so any
physical state is determined by applying a function of the operators
$\hat{a}^+_j$ which commutes with all the constraints $G_a$.
Such a function can  depend only on the operator
$\hat{a}^+_j\hat{a}^+_j$.
Indeed, it must be invariant with respect to the SO(N)-rotations
of the vector $\hat{a}_i^+$. The only independent invariant that can
built of this vector is its square.
Consequently, we find the basis in the physical
subspace
\begin{equation}
\la a^*|n\ra _{ph}=c_n(a^*_ja^*_j)^n,\ \ n=0,1,\ldots\ .
\label{6}
\end{equation}

The normalization factors $c_n$ can be calculated from the equality
$\la n|n'\ra _{ph}=\delta _{nn'}$ and (\ref{5}):
\begin{equation}
c^{-2}_n=\left(\pl/\pl a^*_j\ \pl/\pl a^*_j\right )^n
(a^*_ja^*_j)^n=4^nn!\Gamma (n+N/2)/\Gamma (N/2)\ .
\label{7}
\end{equation}
Non-negative integers $n_i\ (i=1,2,\ldots ,N)$ enumerate the total basis
as the system contains $N$ degrees of freedom, while
the basis (\ref{6}) is
labelled only by one integer $n$, i.e., the system has
only one physical degree of freedom. Note that from the gauge
transformation law (\ref{2}) follows that
the absolute value of the position vector $r=(x^2)^{1/2}\geq 0$
plays the role of a physical variable. We remark that
the phase space spanned by
$r$ and its canonical momentum $p_r$ is a cone
\cite{4}. The fact that the physical configuration (or phase)
space may not coincide with an Euclidean space is usually
ignored in the PI construction for gauge theories. Incidentally, as has
been shown in \cite{5}, it leads to a PI modification, and as a result,
the quasiclassical description can be changed \cite{6}.
For a generic gauge system it is not always possible to
establish the structure of the physical configuration (phase)
space. This problem can be avoided if one uses the PI
suggested below in which unphysical degrees of freedom are not eliminated
explicitly.

Using the Feynman-Kac formula we write the evolution
operator kernel in the physical subspace
\begin{equation}
U^{ph}_t(a^*,a)=\textstyle\sum_{E}^{}\psi ^{ph}_E(a^*)(\psi ^{ph}_E
(a^*))^*e^{-iEt}\ ,
\label{8}
\end{equation}
where $\psi ^{ph}_E(a^*)$ are eigenstates of the Hamiltonian (\ref{3})
satisfying the Dirac condition (\ref{4}).
If in Eq.(\ref{8}) we sum over all eigenstates of
$H$, we get the kernel of the evolution operator $U_t(a^*,a)$ in the total
Hilbert space. Our purpose is to establish a relation between
$U_t$ and $U_t^{ph}$ without
an explicit elimination of unphysical degrees of
freedom by a gauge fixation.

Note that at $t=0$, $U^{ph}_t(a^*,a)=Q(a^*,a)$ is the projector on the
physical subspace, for the functions $\psi ^{ph}_E(a^*)$ compose a
complete orthonormal set.
Note that $H$ and the $G_a$ commute and therefore the total Hilbert
space can be decomposed
into the orthogonal sum of
physical and unphysical subspaces. According to this remark we deduce the
equality
\begin{equation}
U^{ph}_t(a^*,a)=\tint d^N(b^*,b)U_t(a^*,b)Q(b^*,a)\ ,
\label{9}
\end{equation}
i.e., the projection operator $Q$ removes contributions of unphysical
states to the evolution operator. There is a standard representation for
the kernel $U_t(a^*,a)$ by PI \cite{8}
\begin{equation}
U_t(a^*,a)=\int\limits_{}^{}\prod\limits_{\tau =0}^{t}
\frac{d^Na^*d^Na}{(2\pi i)^N}\exp \left[\frac{1}{2}\left(
a^*_j(t)a_j(t)+a^*_j(0)a_j(0)\right)\right]\exp iS\ ,
\label{10}
\end{equation}
where $a^*(t)=a^*,\ a(0)=a$ are the
standard boundary conditions for PI in the
holomorphic representation,
$S=\tint _{0}^{t}d\tau\left[i(a^*_j\dot{a}_j-\dot{a}^*_ja_j)/2-
H(a^*,a)\right]$ is the action of the system including
unphysical degrees of freedom too;
the kernel $H(a^*,a)$
is obtained from the operator $H$ by replacing the operators
$\hat{a}^+_j$ and $\hat{a}_j$ by complex numbers $a^*_j$ and $a_j$,
respectively, after a rearrangement of all $\hat{a}_j$ to the right from
$\hat{a}^+_j$.

Thus, the task is reduced to finding the kernel $Q(a^*,a)$. Since $Q$
is the projector on a physical subspace and the vectors (\ref{6}) form
just another
orthogonal basis in it, we can use the latter to
obtain the resolution of unity in the
physical subspace
\begin{equation}
Q(a^*,a)=\textstyle\sum_{n=0}^{\infty}c^2_n(2\xi)^{2n}=\Gamma (N/2)
\xi ^{1-N/2}I_{N/2-1}(2\xi)\ ,
\label{11}
\end{equation}
where $\xi =1/2(a^*_ja^*_ja_ia_i)^{1/2}$, $I_\nu$ is a modified Bessel
function.

Formulas (\ref{9})-(\ref{11}) solve the above task. The standard form
for $U^{ph}_t$ can also be given:
\begin{equation}
U^{ph}_t(a^*,a)=\int\prod_{\tau}^{}\left(
d^N(a^*,a)\mu (a^*,a)\right)\exp \Phi\exp iS_{ef}\ ;
\label{12}
\end{equation}
here $\mu (a^*,a)$ is some measure in the total phase space of the system,
$S_{ef}$ is an effective action in it, $\Phi$ is a phase associated with a
choice of boundary conditions (cf (\ref{10})). According
to (\ref{8}) the kernel of $U_t^{ph}$ satisfies the equation
$i\pl_tU^{ph}_t(a^*,a)=H(\hat{a}^+,\hat{a}) U^{ph}_t(a^*,a)$ with the
initial condition $U^{ph}_{t=0}=Q$. Note that the kernel (\ref{10})
satisfies the same equation but with the other initial condition:
$U_{t=0}(a^*,a)=\exp\sum a^*_ja_j$.
From this equation we obtain the infinitesimal
kernel of $U^{ph}_\varepsilon$, $\varepsilon \rightarrow 0$,
\begin{eqnarray}
U^{ph}_\varepsilon (a^*,a) &= & Q(a^*,a)\exp \left[-i\varepsilon H_{ef}
(a^*,a)\right]+O(\varepsilon ^2)\ ,
\label{13} \\
H_{ef}(a^*,a) & = & Q^{-1}(a^*,a)H(a^*,\pl/\pl a^*)Q(a^*,a)\ .
\label{14}
\end{eqnarray}
Iterating the kernel (\ref{13}) in accordance with the scalar product
(\ref{5}) we find the path integral representation
of $U^{ph}_t$ for a finite time in the
form (\ref{12}) where
\begin{eqnarray}
\mu(a^*,a)&=&Q(a^*,a)\;; \label{15} \\
S& =&\int\limits_{0}^{t}d\tau\left[\frac{1}{2iQ}\left(\dot{a}^*_j
\frac{\pl}{\pl a^*_j}-\dot{a}_j\frac{\pl}{\pl a_j}\right)Q-H_{ef}
(a^*,a)\right]\;; \label{16} \\
\Phi&=&a^*_j(t)a_j(t)-a^*_j(0)a_j(0)-\frac{1}{2}\ln\frac{Q(a^*(t),a(t))}{
Q(a^*(0),a(0))}\;,\label{17}
\end{eqnarray}
and $a^*(t)=a^*,\ a(0)=a$. Note, if there is no gauge symmetry, then
$Q(a^*,a)=\exp a^*_ja_j$ and Eq.(\ref{12}) turns into (\ref{10}).

Thus, to avoid an explicit elimination of nonphysical
variables in PI, there are
two ways: either to use the projection formula (\ref{9}) or to change both
the measure and action according to formula (\ref{12}),
(\ref{14})-(\ref{17}) in the  ordinary PI over the total phase space.
The main problem in both cases is to find the operator $Q$.

{\bf 3}. Now consider systems with several physical degrees of freedom. Let
us find the operator $Q$ for the Yang-Mills quantum mechanics \cite{7}
with the group SU(2). The model is obtained from Yang-Mills theory
\cite{9} by imposing the condition that all fields depend only on
time, i.e., they are homogeneous in space. The Lagrangian is \cite{10}
\begin{equation}
L={\rm Tr} (\dot{x}-yx)^T(\dot{x}-yx)/2-V(x)\ ;
\label{18}
\end{equation}
here $x$ is a real $3\times 3$ matrix, $y$ is an antisymmetric matrix. If
in the Yang-Mills Lagrangian we identify potentials $A^a_i=A^a_i(t)$ with
$x_{ai}$, where $i,a=1,2,3$ enumerate spatial and isotopic coordinates,
respectively, and $y_{ab}=-g\varepsilon _{abc}A^c_0,\ g$ is a coupling
constant, we get Lagrangian (\ref{18}) in which $V=g^2/4[({\rm Tr}
x^Tx)^2-{\rm Tr}(x^Tx)^2]$, however, our consideration does not depend on
the potential form.

Lagrangian (\ref{18}) is invariant with respect to gauge transformations
of the form
(\ref{2}) where the vector $x$ should be replaced by a matrix $x$ and
$\Omega$ is considered as
an orthogonal $3\times 3$ matrix. The Hamiltonian
formalism for this model is also analogous to that
of the model (\ref{1}). The
momentum canonical conjugated to $y$ vanishes, so we shall not take
this degree of freedom into consideration. The secondary constraints are
generators of isotopic rotations of columns of a matrix $x$. Any real
matrix $x$ can be written in the polar representation $x=u\rho$, where $u$
is an orthogonal matrix and $\rho $ is a positive symmetrical matrix.
Clearly, $u$ contains only unphysical degrees of freedom (they can be
eliminated by the gauge transformation $x\rightarrow u^Tx$). If the PI is
constructed only for physical variables $\rho$
(their number is six because $\rho
=\rho ^T$), the problem of integration over positive definite matrices
arises. It is not equivalent to integration over $\R ^6$ \cite{10}.
Finally, it should be remarked that the physical phase space of the model
differs from the Euclidean space \cite{4}, \cite{5}.
So it is convenient to use the above given
recipe for the gauge-fixing-free PI.

Note that after going over to the holomorphic representation for each
component of the matrix $x_{ai}$, all physical states should be
gauge-invariant $\psi _{ph}(\Omega a^*)=\psi _{ph}(a^*)$, where
$a^*_{aj}=(x_{aj}-ip_{aj})/\sqrt{2}$, $p_{aj}$ are canonical
momenta for $x_{aj}$. One can convince oneself that any vector $\psi
_{ph}(a^*)$ must be a function of the gauge invariant matrix
$(a^{*T}a^*)_{ij}=a^*_{ai}a^*_{aj}$ which describes
six physical degrees of
freedom in this model. So the orthonormal basis in the physical subspace
has the form
\begin{equation}
\la a^*|n\ra =c(n_{ij})[(a^{*T}a^*)_{ij}]^{n_{ij}},\ \ n_{ij}=0,1,
\ldots\ ;
\label{19}
\end{equation}
here $i>j$. The vectors (\ref{19}) are normalized by the scalar product
(\ref{5}) where $N=9$ is a total number of degrees of freedom and $-{\rm
Tr} a^{*T}a=-a^*_{ia}a_{ai}$ is to be placed in the measure in the
exponential argument instead of $-a^*_ja_j$.
The normalization factors
$c(n_{ii})$ (no summation over $i$)  are obtained from (\ref{7})
by setting $n=n_{ii}, N=3$, while to get
$c(n_{ij}),\ i<j$, one should
omit the factor $4^n$  in (\ref{7}) and set
$n=n_{ij},\ N=6$. Now we use again
the resolution of unity in the physical subspace to find $Q$.
A calculation similar to (\ref{11}) yields
\begin{equation}
Q(a^*,a)=\pi ^{3/2}\prod\limits_{i=1}^{3}\xi ^{-1/2}_{ii}I_{1/2}(\xi
_{ii}) \prod\limits_{i<j=1}^{3}\xi ^{-2}_{ij}I_2(2\xi _{ij})\ ,
\label{20}
\end{equation}
where $\xi _{ij}=[(a^{*T}a^*)_{ij}(a^Ta)_{ij}]^{1/2}$. Further by formula
(\ref{14})-(\ref{17}) we restore the physical (gauge-invariant) evolution
operator (\ref{12}) or we can apply (\ref{9}).

{\bf 4}. In
the conclusion we shall show the group method for calculating
the operator $Q$ in any gauge model
with a finite number of degrees of freedom.
Let the brackets $\la\ ,\ \ra$ mean the scalar
product in a representation space of a compact gauge group $G$ and $T_g$
be a group element in this representation. Then
\begin{equation}
Q(a^*,a)=\mu ^{-1}_G\tint d\mu (g)\exp \la a^*_i,T_ga_i\ra\ ;
\label{21}
\end{equation}
here $\mu _G$ is a volume of the group space, $d\mu (g)$ is a right- and
left-invariant Haar measure on $G$, the index $i$ enumerates "particles"
in a representation space, i.e., degrees of freedom are enumerated by $i$
and the group index on which operators $T_g$ act.
The operators $T_g$ are assumed to be
unitary with respect to the scalar product $\la
T_ga^*_i,T_ga_i\ra =\la a^*_i,T_g^+T_ga_i\ra =\la a^*_i,a_i\ra$, i.e.,
$T_g^+=T_{g^{-1}}$. Now we verify easily that $Q(a^*,a)=Q(T_ga^*,T_ga)$.
The latter follows from the unitarity of $T_g$ and
the invariance of the measure
$d\mu (g_1gg_2)=d\mu (g)$. It remains for us to prove the projective
properties of $Q$. After simple calculations we get
\begin{equation}
\tint d^N(b^*,b)Q(a^*,b)\psi (b^*)=\mu ^{-1}_G\tint
d\mu (g)\psi (T^+_ga^*)\ ,
\label{22}
\end{equation}
where $N$ is a total number of degrees of freedom. To derive equality
(\ref{22}), we have used definition (\ref{21}) and the change of
integration variables $b^*_i\rightarrow b^*_i-T^+_ga^*_i$ has been
done. If
$\psi (T_ga^*)=\psi (a^*)$, i.e., it is a physical state, $Q$ acts as
the unit operator because it is a projector on the
physical subspace as follows from Eq.(\ref{22}).
The derivation of PI without gauge
fixation in the Lagrange form will be given elsewhere.

{\bf Acknowledgement.} I am grateful to J.R. Klauder for
useful discussions on the projective method in the path integral
formalism for
constraint systems and his interest in this work as well as for
sending me his papers on the subject \cite{kl}. I wish to thank
the organizers for a financial support of my coming to Dubna.

\end{document}